\documentclass[amsmath,amssymb,reprint,superscriptaddress]{revtex4-1} 

\usepackage{ifthen}		
\newboolean{SubmittedVersion}
\setboolean{SubmittedVersion}{False} 	

\usepackage{graphicx}
\usepackage{amsmath}
\usepackage{amssymb}
\usepackage{amsfonts}
\usepackage{bbold}
\usepackage{dsfont}
\usepackage{textcomp}
\usepackage{upgreek}
\usepackage{color}
\DeclareMathAlphabet{\mathpzc}{OT1}{pzc}{m}{it}

\def\micron{{\ \mu\text{m}}} 

\def\mW{{\ \text{mW}}} 

\def\Hz{{\ \text{Hz}}} 
\def\kHz{{\ \text{kHz}}} 
\def\MHz{{\ \text{MHz}}} 

\def\us{{\ \mu\text{s}}} 
\def\ms{{\ \text{ms}}} 

\def\Rb87{^{87}\text{Rb}} 
\def\Na23{^{23}\text{Na}} 
\def\Li6{^{6}\text{Li}} 

\def\ket#1{\mathinner{|{#1}\rangle}}

{\catcode`\|=\active
 \gdef\Braket#1{\left<\mathcode`\|"8000\let|\BraVert {#1}\right>}}
\def\BraVert{\egroup\,\mid@vertical\,\bgroup}

\newcommand{\expect}[1]{\langle#1\rangle}
\newcommand{\epst}{\tilde\epsilon}
\newcommand{\omx}{\widetilde\omega_x}
\newcommand{\omy}{\widetilde\omega_y}

\newcommand{\ex}{\mathbf{e}_x}
\newcommand{\ey}{\mathbf{e}_y}
\newcommand{\ez}{\mathbf{e}_z}
\newcommand{\exy}{\mathbf{e}_{x,y}}
\newcommand{\exyz}{\mathbf{e}_{x,y,z}}

\newcommand{\ident}{\check{\mathds{1}}}

\begin{document}
\title{Observation of a superfluid Hall effect}
\author{L.~J.~LeBlanc}
\affiliation{Joint Quantum Institute, National Institute of Standards and Technology, and University of Maryland, Gaithersburg, Maryland, 20899, USA}
\author{K.~Jim{\'e}nez-Garc{\'i}a}
\affiliation{Joint Quantum Institute, National Institute of Standards and Technology, and University of Maryland, Gaithersburg, Maryland, 20899, USA}
\affiliation{Departamento de F\'{\i}sica, Centro de Investigaci\'{o}n y Estudios Avanzados del Instituto Polit\'{e}cnico Nacional, M\'{e}xico D.F., 07360, M\'{e}xico}
\author{R.~A.~Williams}
\affiliation{Joint Quantum Institute, National Institute of Standards and Technology, and University of Maryland, Gaithersburg, Maryland, 20899, USA}
\author{M.~C.~Beeler}
\affiliation{Joint Quantum Institute, National Institute of Standards and Technology, and University of Maryland, Gaithersburg, Maryland, 20899, USA}
\author{A.~R.~Perry}
\affiliation{Joint Quantum Institute, National Institute of Standards and Technology, and University of Maryland, Gaithersburg, Maryland, 20899, USA}
\author{W.~D.~Phillips}
\affiliation{Joint Quantum Institute, National Institute of Standards and Technology, and University of Maryland, Gaithersburg, Maryland, 20899, USA}
\author{I.~B.~Spielman}
\affiliation{Joint Quantum Institute, National Institute of Standards and Technology, and University of Maryland, Gaithersburg, Maryland, 20899, USA}

\begin{abstract}{
Measurement techniques based upon the Hall effect~\cite{Hall1879} are invaluable tools in condensed matter physics.
When an electric current flows perpendicular to a magnetic field, a Hall voltage develops in the direction transverse to both the current and the field.  
In semiconductors, this behaviour is routinely used to measure the density and charge of the current carriers (electrons in conduction bands or holes in valence bands) -- internal properties of the system  that are not accessible from measurements of the conventional resistance.  
For strongly interacting electron systems, whose behaviour can be very different from the free electron gas, the Hall effect's sensitivity to internal properties makes it a powerful tool; indeed, the quantum Hall effects~\cite{Klitzing1980,Tsui1982} are named after the tool by which they are most distinctly measured instead of the physics from which the phenomena originate.
Here we report the first observation of a Hall effect in an ultracold gas of neutral atoms, revealed by measuring a Bose-Einstein condensate's transport properties perpendicular to a synthetic magnetic field~\cite{Lin2009b}.
Our observations in this vortex-free superfluid are in good agreement with hydrodynamic predictions~\cite{Recati2001}, demonstrating that the system's global irrotationality influences this superfluid Hall signal.
}\end{abstract}

\maketitle

\begin{figure}[t!]
\begin{center}
\includegraphics[width=89 mm]{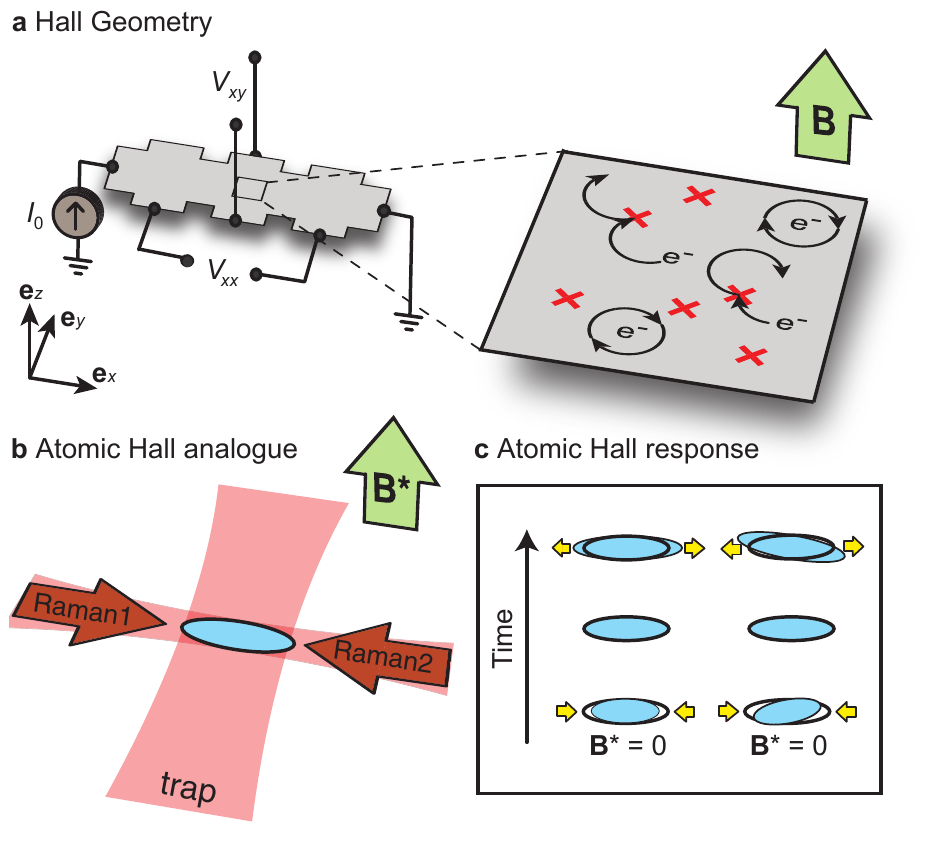}
\end{center}
\caption{\textbf{Hall effect measurements}.  \textbf{a,} 2D electron gas model, pictured as a Hall bar, depicting the transverse (Hall) and longitudinal voltages $V_{xy}$ and $V_{xx}$.  At the microscopic level (right), the electrons move in curved paths under the influence of the magnetic field, and scatter from random impurities (red crosses) with a rate $\tau^{-1}$ (Drude model).  \textbf{b,} The experiment used a 3D BEC, flattened along $\ez$ and  elongated along $\ey$, trapped at the intersection of two 1064~nm laser beams (pink).  A nearly uniform synthetic magnetic field $\mathbf{B}^* = B^* \mathbf{e}_z$ (green arrow) was created via the combination of two Raman laser beams (red arrows) counterpropagating along $\mathbf{e}_x$, and a magnetic field gradient along $\ey$, making a vector potential $\mathbf{A}^* = (-B^*y, 0, 0)$.  \textbf{c,} We generated  a current in the BEC by modulating the amplitude of the external trapping potential along $\mathbf{e}_x$.  The BEC was alternately relaxed (top) and compressed (bottom).  The thick black lines schematically indicate the extent of the BEC in the absence of any current and the blue ellipses indicate the regions occupied by the BEC as the system was modulated.  When $B^*=0$ (left), the transport was primarily along $\mathbf{e}_x$, while when $B^* \neq 0$ (right), the current was deflected along $\mathbf{e}_y$: a superfluid analogue of the Hall effect.} 
\label{fig:setup}
\end{figure}

Microscopically, the Hall effect results from the Lorentz force $\mathbf{F}=q\mathbf{v}\times\mathbf{B}$  experienced by particles with charge $q$ and velocity $\mathbf{v}$ moving in a uniform magnetic field $\mathbf{B}$.  
In the $\ex$-$\ey$ plane perpendicular to $\mathbf{B} = B \ez$, this force acts on a current with density $\mathbf{J} = J_x\ex + J_y\ey$ to generate an electrochemical potential gradient $\boldsymbol{\nabla} V = \check{\boldsymbol{\rho}}_{\rm H} \mathbf{J}$ normal to $\mathbf{J}$, where the Hall part of the 2D resistivity tensor
\begin{align*}
\check{\boldsymbol{\rho}}_{\rm H} &= 
\left(
	\begin{array}{cc}
		0 & -\rho_{xy} \\
		\rho_{xy} & 0
	\end{array}
\right)
\end{align*}
is antisymmetric.  In conventional metals and semiconductors, the Hall resistivity $\rho_{xy} = B/qn(\mathbf{r})$ is related to the carriers' charge $q$ and density $n(\mathbf{r})$, but not to the dissipative resistivity tensor $\check{\boldsymbol{\rho}}_0 = \rho_{xx} \ident$, where $\ident$ is the $2\times2$ identity matrix.  Typically, experiments measure a sample's longitudinal and transverse voltages $V_{xx}$ and $V_{xy}$ (Fig.~\ref{fig:setup}a) from which the resistivity tensor can be inferred~\cite{Simon1999}.  Here, we report an analogous transport measurement of the full resistivity tensor, including the antisymmetric contributions from the Hall effect, in a flattened elongated BEC subjected to a synthetic magnetic field $B^* \mathbf{e}_z$ (in which only the charge-field product~\cite{Spielman2009} $q^*\mathbf{B}^*$ is defined).

The transport characteristics of systems with many particles and sufficiently strong interactions (Coulomb repulsion in electron gases and plasmas, or $s$-wave contact interactions in BECs) resemble those of classical fluids and are described by hydrodynamics.  These hydrodynamics can describe ultracold Bose~\cite{Dalfovo1999} and Fermi~\cite{Giorgini2008} gases, or characterize the collective modes -- plasmons~\cite{Fetter1973} and magnetoplasmons~\cite{DasSarma1982} -- of 2D electron gases (2DEGs).  We show that a BEC in a synthetic magnetic field obeys hydrodynamic equations similar to those describing a 2DEG in a uniform magnetic field.

 In a simple Drude model~\cite{AshcroftBook1976} description of a 2DEG in a magnetic field (Fig.~\ref{fig:setup}a), scattering from impurities gives rise to resistance while a position-dependent potential $V(\mathbf{r})$ controls the electron density.  The collective electron dynamics can be expressed in terms of hydrodynamic continuity and Euler equations:
\begin{align}
0 &= \partial_t n(\mathbf{r}) + \frac{1}{m} \boldsymbol{\nabla} \cdot \left[\mathbf{p}_{\rm m}(\mathbf{r}) n(\mathbf{r})\right]\label{eq:DrudeContinuity}\\
0 &=\frac{D \mathbf{p}_{\rm m}(\mathbf{r})}{Dt}+ \left(\begin{array}{cc}
							\tau^{-1} &\text{--}\Omega_\mathrm{C} \\
							\Omega_\mathrm{C} & \tau^{-1}
							\end{array}\right) \mathbf{p}_{\rm m}(\mathbf{r}) + \boldsymbol{\nabla}\left[U(\mathbf{r}) + gn(\mathbf{r})\right],		\label{eq:DrudeEuler}
\end{align}
where $\mathbf{r} = (x,y)$; $\mathbf{p}_{\rm m} = [m/qn(\mathbf{r})]\mathbf{J}$ is the mechanical momentum of particles with mass $m$, charge $q$, and density $n(\mathbf{r})$; $\Omega_{\rm C} = qB/m$ is the cyclotron frequency for magnetic field strength $B$;   $U(\mathbf{r}) = qV(\mathbf{r})$ is the potential energy;   $g = 2\pi\hbar^2/m$ accounts for the effects of Fermi pressure in this noninteracting 2DEG~\cite{Fetter1973};  $\tau$ is the momentum relaxation time due to scattering from impurities; and $D/Dt = \partial_t + m^{-1}[\mathbf{p}_{\rm m}(\mathbf{r}) \cdot \boldsymbol{\nabla}]$ is the convective derivative.  The matrix in equation~\eqref{eq:DrudeEuler} is proportional to the resistivity tensor {\color{black} [through the factor $m/q^2n(\mathbf{r})$]} and has equal diagonal components from $\check{\boldsymbol{\rho}}_0$ (conventional resistance) and antisymmetric off-diagonal components from $\check{\boldsymbol{\rho}}_{\rm H}$ (Hall resistance).  
To compare trapped interacting BECs with the 2DEGs described above, the superfluid hydrodynamic (SFHD) equations may be extended~\cite{Recati2001, Cozzini2003a} to include a time-independent artificial magnetic field $\mathbf{B}^* = B^* \ez$.  The resulting 2D continuity and Euler equations are identical to equations \eqref{eq:DrudeContinuity} and \eqref{eq:DrudeEuler}, complete with the off-diagonal antisymmetric components of the resistivity tensor -- the Hall term -- proportional to the  cyclotron frequency $\Omega_{\rm C} = q^*B^*/m$ (see Supplementary information).  The BEC's trap provides the external potential energy $U(\mathbf{r}) = m(\omega_x^2x^2 + \omega_y^2 y^2)/2$ with trap frequencies $\omega_{x,y}$ along $\exy$.  
The interaction term $gn(\mathbf{r})$ accounts for the mean-field energy due to contact interactions in the BEC.
 In principle, the resistivity tensor's diagonal components are zero due to the BEC's superfluidity; however,  atom loss can imitate damping and we  retain a phenomenological dissipative term proportional to $\tau^{-1}$.  Though the systems with which we work are fully three-dimensional, the dynamics of interest are in the plane perpendicular to $q^*\mathbf{B}^*$ and are described by these 2D equations.  For quantitative comparisons to data, calculations were performed using the 3D version of equations~\eqref{eq:DrudeContinuity} and~\eqref{eq:DrudeEuler} (see Supplementary information).  

By expressing the SFHD equations in terms of the gauge-invariant mechanical momentum $\mathbf{p}_{\rm m}$, which describes the actual motion of particles, we depart from the conventional expressions in terms of the canonical momentum $\mathbf{p} = \mathbf{p}_{\rm m} + q^*\mathbf{A}^*$ (where $\mathbf{A}^*$ is the artificial vector potential such that $\mathbf{B}^* = \boldsymbol{\nabla} \times \mathbf{A}^*$), which is directly related to the gradient of the superfluid order parameter's phase.  The distinction between $\mathbf{p}$ and $\mathbf{p}_{\rm m}$ has tangible consequences when considering the condition of irrotationality $\nabla \times \mathbf{p} = 0$.  When a synthetic magnetic field is present, irrotationality requires equilibrium fluid flow, with $\nabla \times \mathbf{p}_{\rm m} = -q^*\mathbf{B}^*$.  
This need not be the case at places like vortex cores, where superfluid density is zero; however, in our trapped BEC, vortices are energetically allowed only when $\Omega_{\rm C}$ is greater than a critical value~\cite{Lundh1997,Lin2009b}. Unless noted, we remained in the irrotational regime where  $\Omega_{\rm C}$ is below this critical value.

In this experiment, we studied mass transport in a BEC subject to a synthetic magnetic field $\mathbf{B}^*$ created using Raman dressing~\cite{Spielman2009, Lin2009b}.  While previous experiments used rotating traps to exploit the equivalence between the Coriolis and Lorentz forces~\cite{Fetter2009}, methods that use atom-light coupling~\cite{Dalibard2011} instead of rotation allow for greater geometric freedom and potentially increased field strengths~\cite{Cooper2011}.  The Raman technique used here enabled our experiment's highly anisotropic BECs, whose 4:1 aspect ratio (Fig.~\ref{fig:setup}b) mimicked a typical bar-like Hall geometry and facilitated  identification of the Hall response.  

Our experiments began with a nearly-pure $\Rb87$ BEC of about $2\times 10^5$ atoms in the $\ket{F = 1, m_F = -1}$ internal state~\cite{Lin2009} confined in a crossed optical dipole trap with frequencies $(\omega_x, \omega_y, \omega_z)/2\pi \approx (11.3, 42.5, 90)$~Hz (Fig.~\ref{fig:setup}b).  
Laser beams counterpropagating along $\ex$ Raman-coupled spin states $m_{F=1} = 0,\pm 1$ of differing momentum to create new dressed eigenstates -- spin-momentum superpositions~\cite{Lin2009b}. 
 The dressed atoms experienced a synthetic vector potential~\cite{Lin2009b}, which in the presence of an appropriate real magnetic field gradient  gave  $q^*\mathbf{A}^* = (-q^*B^*y,0,0)$ and the associated synthetic charge-field product $q^*\mathbf{B}^* = q^*{B}^*\ez$.

\begin{figure}[t!]
\begin{center}
\ifthenelse{\boolean{SubmittedVersion}}{}{\includegraphics[width=89 mm]{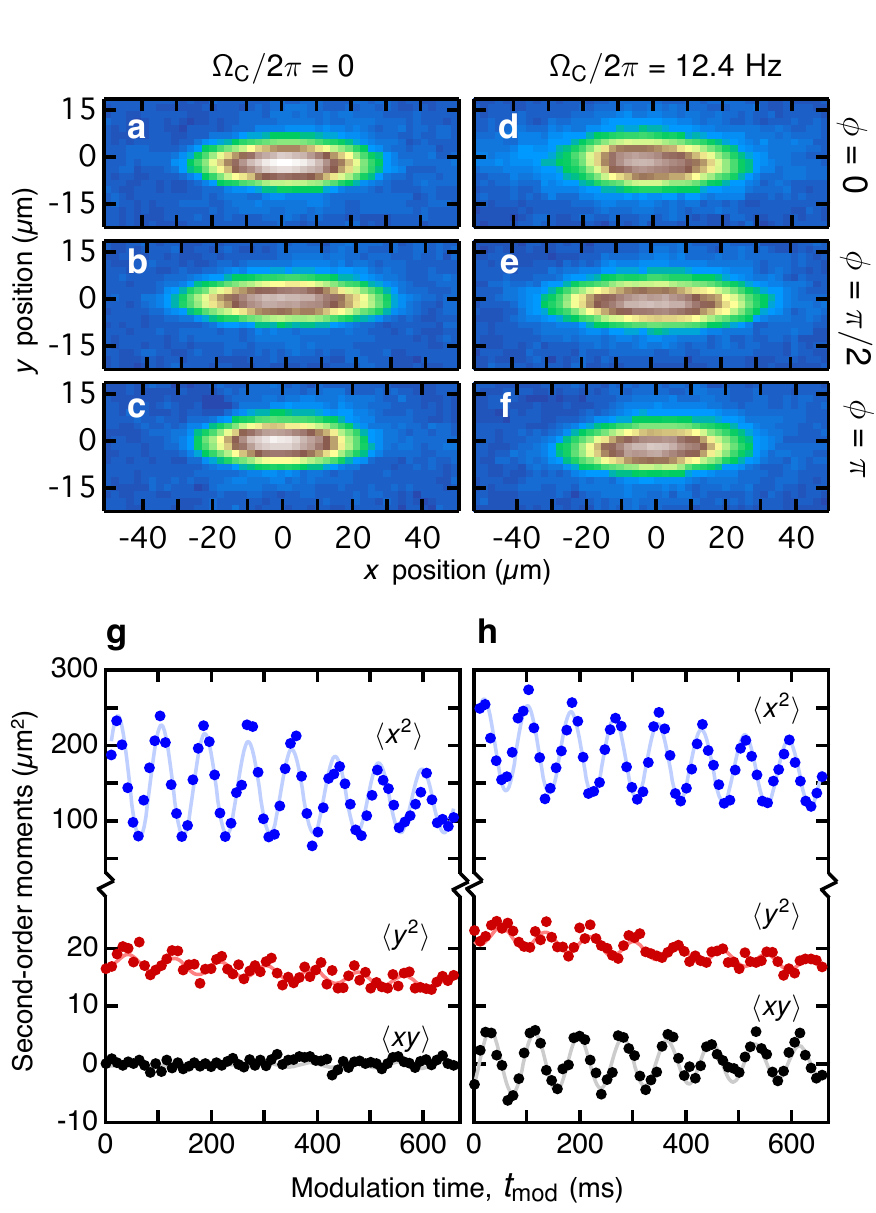}}
\end{center}
\caption{\textbf{Time-dependent Hall signal.} \textbf{a-f} \emph{In situ} absorption images reveal the atomic density distribution $n(x,y)$ at different times during the 83.3~ms modulation cycle: \textbf{a,d} $t = $157.3~ms, \textbf{b,e} $t = $178.1~ms, \textbf{c,f} $t = $198.9~ms.  For $\Omega_{\rm C} = 0$ (\textbf{a-c}), the density modulation is primarily along $\ex$, while for $\Omega_{\rm C}/2\pi = 12.4$~Hz (\textbf{d-f}) there is a visible skew to the cloud, which changes as a function of modulation time.    \textbf{g,h} The evolution of the second-order moments of $n(x,y)$ [equation \eqref{eq:xy}] during the steady-state modulation of the trapping potential, without (\textbf{g}) and with (\textbf{h}) synthetic magnetic field.  Pale curves are guides to the eye.}
\label{fig:MeasureHall}
\end{figure}

To measure transport, we  generated a mass current  (a velocity field) in the BEC by perturbatively modulating the trap along $\ex$ (Fig.~\ref{fig:setup}c). We modified the potential energy $U(x,y,z) = (\kappa_x x^2 +\kappa_y y^2 + \kappa_z z^2)/2$, where $\kappa_i = m_i\omega_i^2$ is the spring constant along $\mathbf{e}_i$, by adding a time-dependent drive $\delta U_x = (\delta \kappa_x x^2/2) \sin(\omega t)$ with amplitude $\delta \kappa_x$ and drive frequency $\omega$.  To ensure linear response, we chose $\delta \kappa_x \leq 0.13 \kappa_x$; to avoid transients, we smoothly increased $\delta \kappa_x$ from zero in $500\ms$.  After driving in steady-state for a time $t_{\rm mod}$, we measured the atomic density, either \emph{in situ} or after time-of-flight (TOF).  We characterized the system's temporal response to the drive by repeating this procedure for 64 equally spaced values of $t_{\rm mod}$, the longest of which encompassed eight cycles of modulation.
 
When the geometry of the trap deformation is well-overlapped with one of the system's low-energy collective excitations~\cite{Stringari1996a,Murray2007} -- eigenmodes of the SFHD equations -- the response is enhanced as $\omega$ approaches that mode's resonance frequency.  Such collective modes have been measured in both Bose and Fermi gases~\cite{Dalfovo1999,Giorgini2008}.  In this work, we predominantly drove the quadrupole-like ``$X_2$'' eigenmode  (whose frequency is interaction-dependent) corresponding to compression and expansion primarily along $\ex$ (an oscillatory $\expect{x^2}$ response), with smaller out-of-phase contributions, due to interactions, along $\ey$ and $\ez$ ($\expect{y^2}$ and $\expect{z^2}$ responses).  When the synthetic magnetic field is non-zero,  correlated transport along $\ex$ and $\ey$ appears (an $\expect{xy}$ response) resulting from mixing with the scissors mode~\cite{Marago2000}.  This $B^*$-dependent correlated transport is the Hall effect.  Had we excited the ``$X_1$'' dipole (sloshing) mode, we would have observed the Hall effect, but would not have probed internal properties related to interactions, as described by Kohn's theorem~\cite{Kohn1961}.

We absorption-imaged the time-dependent column density distributions $n(x,y)$ either \emph{in situ}, to directly determine the density response, or after TOF, to determine the momentum response.  For \emph{in situ} measurements, we imaged the atoms immediately following release from the trap ($t_{\rm off}< 1 \us$).  For TOF measurements, we allowed the atoms to expand freely for 36.2~ms before imaging.  During the first 2~ms of TOF, we deloaded~\cite{Williams2012} the Raman-dressed atoms into the $\ket{F=1, m_F = 1}$ bare state before removing the Raman dressing fields.  

These density distributions offer snapshots of the dynamically evolving BEC at $t_{\rm mod}$.  For each $t_{\rm mod}$, we extract three independent second-order moments of the density distribution, $\expect{x^2}$, $\expect{y^2}$, and $\expect{xy}$:
\begin{align}
\expect{x_i x_j} = \frac{\sum x_i x_j ~{n}(x,y)}{\sum{n}(x,y)},\label{eq:xy}
\end{align}
where  $x_i \in \{x, y\}$ was measured from the centre of the density distribution $(\expect{x} = \expect{y} = 0)$.  

\begin{figure}[t!]
\begin{center}
\ifthenelse{\boolean{SubmittedVersion}}{}{\includegraphics[width=89 mm]{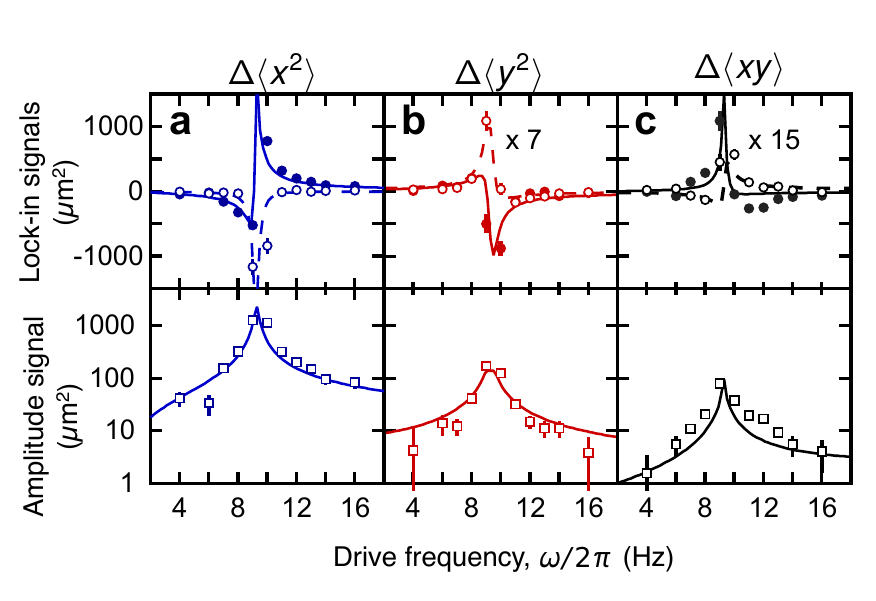}}
\end{center}
\caption{\textbf{Spectra of response to trap modulation}.  These responses, measured after TOF, are dominated by the \emph{in situ} momentum.  Second order moment amplitudes $\Delta\expect{x^2}$, $\Delta\expect{y^2}$, and $\Delta\expect{xy}$, as a function of modulation frequency $\omega$ (\textbf{a}, \textbf{b}, and \textbf{c}, respectively).  ~\textbf{Top row}, In- and out-of-phase responses (open and closed symbols, respectively), as determined by a lock-in analysis (see Supplementary information).  Dashed (solid) curves show the results of a SFHD fit to the response for the in-phase (out-of-phase)  components.  The fit was performed simultaneously to all six sets of data.  The uncertainties  reflect the standard deviation of the mean.  \textbf{Bottom row}, Amplitude response (root-sum-square of the above results).  Solid curves show the calculated response using the above fit results.  }
\label{fig:quadandamp}
\end{figure}

Figure \ref{fig:MeasureHall} shows ${n}(x,y)$ at three points during the modulation cycle  without (Figs.~\ref{fig:MeasureHall}a-c) and with (Figs.~\ref{fig:MeasureHall}d-f) the synthetic magnetic field. While the compression of the cloud along the drive direction $\ex$ was always the most evident signal, a nonzero synthetic magnetic field ($\Omega_{\rm C} \neq 0$) tilted the clouds as the velocity field responded to the drive current,  macroscopically manifesting the BEC's superfluid Hall resistivity.  In Figs.~\ref{fig:MeasureHall}d-f, the Hall signal $\expect{xy}$ oscillates about zero and reaches its minimum, zero, and maximum, respectively.  The temporal responses of the second-order moments are plotted in Figs.~\ref{fig:MeasureHall}g ($\Omega_{\rm C} = 0$) and \ref{fig:MeasureHall}h ($\Omega_{\rm C} \neq 0$) for $\omega/2\pi = 12\Hz$.  This modulation frequency is above resonance (Fig.~\ref{fig:quadandamp}): close enough for a strong signal, but  far enough to spectrally resolve the natural mode from the driven mode.  While $\expect{x^2}$ and $\expect{y^2}$ were not significantly affected with the introduction of the synthetic field,  a weak $\Omega_{\rm C}$-dependence of the parallel $\expect{x^2}$ and perpendicular $\expect{y^2}$ moments arises from variations in the effective trapping frequencies as field strength increases~\cite{Cozzini2003a,Spielman2009}.  

\begin{figure}[t!]
\begin{center}
\ifthenelse{\boolean{SubmittedVersion}}{}{\includegraphics[width=89 mm]{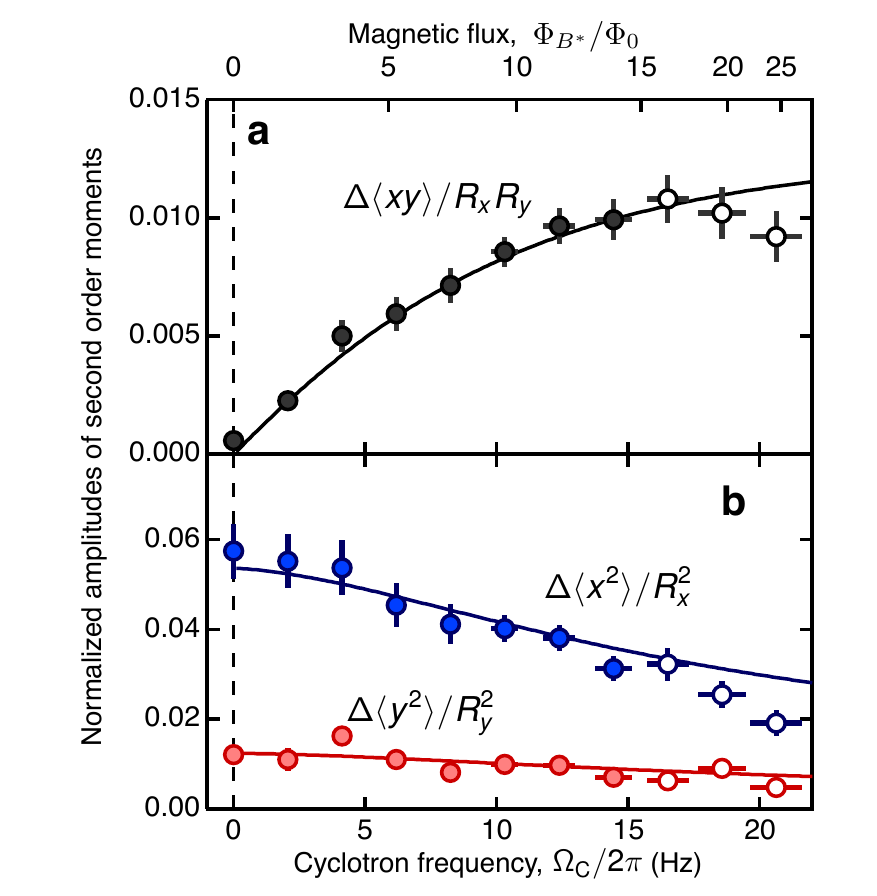}}
\end{center}
\caption{\textbf{Hall effect}.  \textbf{a,} Measurements of the Hall signal $\expect{xy}/R_x R_y$  (black)  for modulation frequency $\omega/2\pi = 12\Hz$ at various synthetic magnetic field values, from \emph{in situ} images.  Thomas-Fermi radii  $R_x, R_y$ are from fits to the data.  Filled circles indicate data in which the system has no vortices and open circles indicate data for which TOF images indicate that vortices were likely present.  The top axis indicates the calculated ``magnetic flux'' penetrating the BEC~\cite{Lin2009b}, using $\Phi_{B^*}/\Phi_0 = \Omega_{\rm C} m\mathcal{A}/h$, where $\mathcal{A} = \pi R_xR_y$ is the area encompassed by the BEC in the $\ex$-$\ey$ plane, which increases with $\Omega_{\rm C}$.   The vertical uncertainties reflect the standard deviation of the mean; the horizontal uncertainties reflect only the experimental uncertainty in $\Omega_{\rm C}$ and not the Thomas-Fermi radii.  \textbf{b,} The synthetic field dependence of the time-dependent part of the normalized second order moments $\Delta \expect{x^2}/R_x^2$ (blue circles) and $\Delta \expect{y^2}/R_y^2$ (red circles).  The solid curves show the calculated SFHD response, which was simultaneously fit to all three moments.  The signals' nonlinearity is due to the decreasing  collective mode resonance frequency with increasing $\Omega_{\rm C}$.}
\label{fig:OmegaC}
\end{figure}

Figure \ref{fig:quadandamp} shows the spectrum of the response across the low-frequency $X_2$ resonance. 
We measured the time-dependent second-order moments $\expect{x^2}$, $\expect{y^2}$, and $\expect{xy}$ and used a lock-in technique to determine the response amplitudes $\Delta{\expect{x^2}}$, $\Delta{\expect{y^2}}$, and $\Delta{\expect{xy}}$ for each moment  both in-phase and out-of-phase with the drive.
We used the TOF method to ensure a strong signal even far from resonance.  At a moderate synthetic magnetic field strength [$\Omega_{\rm C} /2\pi =10.5(5)\Hz$], no vortices entered and equations \eqref{eq:DrudeContinuity}, \eqref{eq:DrudeEuler} apply.  We simultaneously fit all six $\omega$-dependent responses to the solutions of the linearized SFHD equations (propagated through TOF without linearization, see Supplementary information) and extracted three free parameters: the phenomenological damping term $\tau^{-1} =4.4(4)~\text{s}^{-1}$, which is the diagonal part of the resistivity matrix in equation \eqref{eq:DrudeEuler}; the effective trap frequency, $\omega_x/2\pi =6.07 (2)$~Hz; and the drive strength $\delta\kappa_x/\kappa_x = 0.074(2)$.  The latter two parameters are in reasonable agreement with the directly measured quantities $\omega_x/2\pi =6.1(2)$ Hz and $\delta\kappa_x/\kappa_x = 0.10(1) $.

While the above data are well-characterized by and are in good agreement with a SFHD description and its TOF propagation, \emph{in situ} measurements provide a clearer demonstration of Hall physics.  For different values of $\Omega_{\rm C}$, we measured the BECs' time-dependent density profiles and calculated normalized second-order moments $\expect{x^2}/R_x^2$, $\expect{y^2}/R_y^2$, and $\expect{xy}/R_xR_y$ from  fits to skewed Thomas-Fermi distributions with radii $R_x$ and $R_y$ (see Supplementary information).  As in Fig.~\ref{fig:MeasureHall}, the modulation drive frequency was $\omega/2\pi = 12$~Hz.

Figure \ref{fig:OmegaC} shows the amplitude response of the normalized second order moments along with simultaneous fits to the linear-response predictions of SFHD using a single fit parameter, $\delta\kappa_x$ (with the previously measured $\omega_x$ and $\tau^{-1}$ as fixed parameters).  For small $\Omega_{\rm C}$, we find good agreement between our measurements and SFHD.   The nonlinearity of $\Delta\expect{xy}/R_xR_y$ at high fields is caused by the shift of the $X_2$ resonance away from the fixed drive frequency $\omega$ as $\Omega_{\rm C}$ increases~\cite{Cozzini2003a,Spielman2009}.  {\color{black} For small $\Omega_{\rm C}$, where the resonance is nearly constant, the slope of $\Delta\expect{xy}/R_xR_y$ vs $\Omega_{\rm C}$ is a measure of the Hall resistivity $\rho_{xy} = m\Omega_{\rm C}/q^2 n$ averaged over the cloud's changing density.} The good agreement between our measurements and the irrotational SFHD model demonstrates the utility of this new Hall technique as a simple, quantitative way to probe microscopic properties, like the superfluid Hall resistivity, of a quantum degenerate gas.  
Beyond $\Omega_{\rm C}/2\pi\approx 15\Hz$, indications of vortices were evident in TOF; therefore, we expect that the predictions of vortex-free SFHD no longer apply.  In the large-$B^*$ limit of ``diffused vorticity,''~\cite{Cozzini2003} in which many vortices have entered the BEC and ``ordinary'' fluid behaviour with $\nabla \times \mathbf{p}(\mathbf{r}) =  q^* \mathbf{B}^*$ is valid,  calculations show that the superfluid Hall signal approaches zero, but does not become negative as in some superconducting materials~\cite{Smith1997,Huber2011}.

Inspired by condensed-matter techniques, we have demonstrated Hall physics in a BEC and showed that it is sensitive to the properties of an interacting, irrotational BEC.  This  macroscopic Hall measurement technique can be extended to probe the microscopic properties of more complicated ultracold configurations, such as 2D systems with vortices (arising from thermal fluctuations~\cite{Hadzibabic2006} or synthetic magnetic fields~\cite{Lin2009b}), systems with spin-orbit coupling~\cite{Bijl2011}, and quantum Hall systems~\cite{Sorensen2005}.


\vspace{20pt}
\noindent{\bf Acknowledgements} We appreciate conversations with J.~V.~Porto.  This work was partially supported by ONR; ARO with funds from both DARPA's OLE program and the Atomtronics MURI; and the NSF through the Physics Frontier Center at JQI.  L.~J.~L.  acknowledges support from NSERC; K.~J.-G.  acknowledges CONACYT; and M.~C.~B.  acknowledges NIST-ARRA.
\\
\\
\noindent{\bf Author Contributions} All authors contributed to writing of the manuscript.  L.~J.~L.  executed the final calculations, and led the data taking effort in which the remaining authors participated.  W.~D.~P.  and I.~B.~S.  supervised this work and I.~B.~S.  proposed the initial measurements in accordance to preliminary numerical and analytic calculations.
\\
\\
\noindent{\bf Author Information} Correspondence and requests for materials
should be addressed to I.~B.~S.  (ian.spielman@nist.gov).

\clearpage
\section*{Supplementary information}

\setcounter{equation}{0}
\renewcommand{\theequation}{M.\arabic{equation}}

\subsection*{Drude model hydrodynamics}
In a 2DEG, the equation of motion for the electrons  is given by Euler's equation~\cite{Fetter1973,DasSarma1982}
\begin{align*}
\frac{D \mathbf{p}_{\rm m}(\mathbf{r},t)}{D t} + \frac{1}{\tau} \mathbf{p}_{\rm m}(\mathbf{r},t) +\frac{q}{m} \mathbf{B}\times \mathbf{p}_{\rm m}(\mathbf{r},t) \nonumber\\+\boldsymbol\nabla U(\mathbf{r}) +\frac{1}{n}\boldsymbol\nabla P(\mathbf{r})= 0,
\end{align*} 
where $\mathbf{p}_{\rm m}(\mathbf{r},t) = m~d\mathbf{r} / dt$ is the electrons' mechanical momentum, $q$ their charge, $\tau$ the mean time between scattering events, $U(\mathbf{r})=qV(\mathbf{r})$ the external potential energy, and $P(\mathbf{r})$  the local pressure.  This hydrodynamic description involves the momentum distribution $\mathbf{p}_{\rm m}(\mathbf{r},t)$ of the fluid's velocity field, rather than the momenta of individual particles, requiring the use of the convective derivative $D/Dt =\partial_t + m^{-1} [ \mathbf{p}_{\rm m} \cdot \nabla]$.
For a noninteracting single-component 2DEG, the internal energy in an area $\mathcal{A}$ is $U_{\rm int} = \pi\hbar^2 N^2/m\mathcal{A} = N E_{\rm F} /2$, where $N$ is the number of electrons and $E_{\rm F}$ the Fermi energy.   From the energy, the local pressure $P = -(\partial U_{\rm int}/\partial \mathcal{A})_{N} = \pi \hbar^2 n^2/m$ can be determined.  Because $P \propto n^2$, the force due to the internal Fermi pressure is proportional to the gradient of the density, and is (coincidentally) directly analogous to the effect of contact interactions in a BEC.  

\subsection*{Superfluid hydrodynamics}
The SFHD equations described in the text  were derived from the time-dependent Gross-Pitaevskii equation (GPE)~\cite{PethickSmithBook}
\begin{align*}
i\hbar \frac{\partial \psi(\mathbf{r}, t)}{\partial t} = -\check{\mathbf{m}}^{-1} \frac{\hbar^2}{2} \boldsymbol\nabla^2 \psi(\mathbf{r}, t) + V(\mathbf{r})  \psi(\mathbf{r}, t) \nonumber\\+ U(\mathbf{r}) |\psi(\mathbf{r}, t))|^2\psi(\mathbf{r}, t)
\end{align*}
with the order parameter (or wavefunction) $\psi(\mathbf{r},t)$ and inverse mass tensor $\check{\mathbf{m}}^{-1}$ (which may in general be anisotropic).  By writing the order parameter  $\psi(\mathbf{r},t) = \sqrt{n(\mathbf{r},t)} e^{i\nu(\mathbf{r},t)}e^{-i q^*\int \mathbf{A}^*(\mathbf{r},t)\cdot d\mathbf{s}/ \hbar}$ in terms   of the density  $n(\mathbf{r},t)$, the phase $\nu(\mathbf{r},t)$, and the vector potential $q^* \mathbf{A}^*(\mathbf{r},t)$ (where the integral is evaluated over the position of the particle $\mathbf{s}$), and neglecting the quantum pressure term~\cite{PethickSmithBook}, the GPE can be reformulated as the  continuity and Euler equations:
 \begin{align}
0 =&\ \partial_t n(\mathbf{r},t) + \check{\mathbf{m}}^{-1}\boldsymbol{\nabla} \cdot \left[\mathbf{p}_{\rm m}(\mathbf{r},t) n(\mathbf{r})\right]\label{eq:continuityMethods}\\
0 =&\ \partial_t \mathbf{p}_{\rm m}(\mathbf{r},t) + \boldsymbol{\nabla}\left[\tfrac{1}{2}\mathbf{p}_{\rm m}(\mathbf{r},t) \cdot \check{\mathbf{m}}^{-1}\cdot \mathbf{p}_{\rm m}(\mathbf{r},t) + U(\mathbf{r},t)\right.\nonumber\\ & \left.   \qquad\qquad\qquad\qquad+gn(\mathbf{r})\right] \label{eq:EulerMethods}+ \tau^{-1} \mathbf{p}_{\rm m}(\mathbf{r},t),
\end{align}
where $ \mathbf{p}_{\rm m}(\mathbf{r},t) = \mathbf{p} (\mathbf{r},t)- q^* \mathbf{A}^*(\mathbf{r})$ is the mechanical momentum and $\mathbf{p}(\mathbf{r},t) = \hbar \boldsymbol \nabla \nu(\mathbf{r},t)$ is the canonical momentum.  A phenomenological damping constant  $\tau$ is also included, which is attributed to loss of atoms from the condensate due to photon scattering from the Raman beams.  It can be shown~\cite{Nikuni1999}  that such loss, through exchange of atoms with the thermal cloud, results in effective momentum damping within the linear approximation (below) 

In this context, the important property of superfluidity is its irrotationality, and this condition, $\boldsymbol{\nabla} \times \mathbf{p}(\mathbf{r},t)  = 0$, places a further constraint upon the dynamics.   When  $q^*B^*$ is  sufficiently small~\cite{Lundh1997,Lin2009b} and the Landau gauge vector potential $\mathbf{A}^*(\mathbf{r}) = (-B^* y, 0, 0)$ is assumed, the equilibrium condition (determined by setting the time derivatives to zero  in equations (\ref{eq:continuityMethods}) and
(\ref{eq:EulerMethods})) is found when $\mathbf{p} = \tfrac{1}{2}q^*B^*(1+{\epst})\boldsymbol{\nabla}(xy)$.  The solutions of equations \eqref{eq:continuityMethods} and \eqref{eq:EulerMethods} in equilibrium yield the time-independent \emph{in situ} density distribution of the BEC, which is of the usual Thomas-Fermi form~\cite{PethickSmithBook}, but with effective trapping frequencies $\omx^2 = \omega_x^2 + \tfrac{1}{2}\Omega_{\rm C}(1+\epst)$, $\omy^2 = \omega_y^2 + \tfrac{1}{2}\Omega_{\rm C}(1-\epst)$.  The anisotropy parameter $\epst = (\omx^2-\omy^2)/(\omx^2+\omy^2)$ is determined implicitly~\cite{Cozzini2003a} for each $\Omega_{\rm C}$.  The $\Omega_{\rm C}$-dependence of these effective trapping frequencies $\omx$ and $\omy$ shifts the resonance frequencies of the effective modes as a function of synthetic magnetic field strength $q^*B^*$.  

 As written, equations \eqref{eq:continuityMethods} and \eqref{eq:EulerMethods}
 apply in any dimension.  Assuming 2D, a diagonal uniform mass tensor ($\check{\mathbf{m}} = m \ident$)  and the spatially dependent 2D interaction term, equations  \eqref{eq:continuityMethods} and \eqref{eq:EulerMethods} are transformed using standard vector identities to equations (1) and (2).  In order to compare calculations using the SFHD equations to our data, we used the three-dimensional (3D) form, with the mean-field interaction parameter $g = 4 \pi \hbar^2 a_{\rm s} /m$ (where $a_{\rm s}$ is the s-wave scattering length), and the anisotropic effective mass tensor
\begin{align*}
\check{\mathbf{m}}^{-1} = \left(\begin{array}{ccc} 1/m^*_x & 0 & 0 \\ 0 & 1/m & 0 \\ 0 & 0 & 1/m\end{array} \right), 
\end{align*}
where $m^*_x$ is the effective mass calculated from the modified dispersion relation along $\ex$, which arises from Raman dressing~\cite{Spielman2009}.

%

To calculate the dynamics under a time-periodic modulation of the trapping potential, we assumed a perturbative change $\delta U_x$ of the trapping potential energy $U$ as described in the main text, and for convenience used a complex time dependence: $U(\mathbf{r},t) = U(\mathbf{r}) + \delta U_x(\mathbf{r})e^{-i\omega t}$.  Under this approximation, the linearization of equations \eqref{eq:continuityMethods} and \eqref{eq:EulerMethods} gives
\begin{align}
0 =& \partial_t\delta n(\mathbf{r},t) + \check{\mathbf{m}}^{-1} \boldsymbol{\nabla} \left[n_0(\mathbf{r}) \delta \mathbf{p}(\mathbf{r},t)+\right.\nonumber\\&\qquad\qquad\qquad \left.(\mathbf{p}_0(\mathbf{r})-q^*\mathbf{A}^*(\mathbf{r}))\delta n(\mathbf{r},t)\right]\label{eq:linContinuity}\\
0 =&\partial_t\delta \mathbf{p}(\mathbf{r},t) + \boldsymbol{\nabla} \left[(\mathbf{p}_0(\mathbf{r}) - q^*\mathbf{A}^*(\mathbf{r}))\cdot \check{\mathbf{m}}^{-1}\cdot\delta \mathbf{p}(\mathbf{r},t) \right. \nonumber\\ &\left. \qquad\qquad+ \delta U (\mathbf{r})+g\delta n(\mathbf{r},t)\right]+\tau^{-1} \delta \mathbf{p}(\mathbf{r},t) \label{eq:linEuler},
\end{align} 
where we modelled the time-dependent density and momentum as the equilibrium distributions plus a small contribution oscillating at the drive frequency: $n(\mathbf{r},t) = n_0(\mathbf{r}) + \delta n(\mathbf{r},t)$ and $\mathbf{p}(\mathbf{r},t) = \mathbf{p}_0(\mathbf{r}) + \delta \mathbf{p}(\mathbf{r},t)$. (Note that for our experiments $\partial_t\mathbf{A}^*(\mathbf{r},t) = 0$.)   In the 3D calculations, the density modulation was assumed to be of the form $\delta n(\mathbf{r},t) = e^{-i\omega t} (\delta n_0 + \delta n_x x^2 + \delta n_y y^2 + \delta n_z z^2 + \delta n_{xy} xy)$ and the momentum modulation $\delta \mathbf{p}(\mathbf{r},t) = e^{-i \omega t} \nabla\delta\nu(\mathbf{r})$ where $\delta\nu(\mathbf{r}) = \delta \nu_0 + \delta \nu_x x^2 + \delta \nu_y y^2 +\delta \nu_z z^2 + \delta \nu_{xy} xy$ [which ensures irrotationality: $\nabla \times \delta\mathbf{p}(\mathbf{r},t) = 0$].  When the synthetic field is present, non-zero coefficients of the $xy$ terms are required to obtain a solution and indicate a Hall response.  
To extract the temporal dependence of $\delta n(\mathbf{r},t)$ and $\delta p(\mathbf{r},t)$, we calculated the responses at equally spaced $t \in [0 , 2\pi/\omega]$, taking the imaginary part of $[-\delta n(\mathbf{r},t)]$ and $[-\delta \mathbf{p}(\mathbf{r},t)]$ at each $t$ to obtain the response to the $\sin(\omega t)$ drive.  

To compare these results with our TOF measurements, we propagated the calculated \emph{in situ} density and momentum distributions through 36.2~ms TOF~\cite{Castin1996,Dalfovo1999} using the SFHD equations \eqref{eq:continuityMethods} and \eqref{eq:EulerMethods} with $\mathbf{A}^*$, $U$ and $\tau^{-1}$ set to zero.  We projected these numerical solutions for the density profile into a two-dimensional plane to mimic absorption imaging and analyzed them in the same manner as the data.

This theoretical model is fit to the experimentally determined second-order moments (Fig.~3).  
As an additional check, we compared the experimental results  determined by direct integration with results obtained by fitting Thomas-Fermi profiles to  the density distributions $n(x,y)$ and comparing the fit parameters to those expected from the SFHD model.  Both methods of analysis were in excellent agreement with the model, and each other.

\subsection*{System preparation}
After standard laser cooling and trapping procedures~\cite{Lin2009}, our experiments began with nearly pure $\Rb87$ BECs of approximately $2\times 10^5$ atoms in the $\ket{F=1,m_F=-1}$ state.  The condensate was trapped at the intersection of two 1064~nm laser beams, propagating along $\mathbf{e}_{x}$ (with $\approx\!500\mW$ focussed to $1/e^2$ radii of $w_{y}\approx 120\micron$ and $w_{z}\approx 50\micron$) and $\mathbf{e}_{y}$ ($\approx\!125\mW$  with $w_{x}\approx w_{z}\approx 160\micron$), crossing near their focii.  

We implemented a synthetic vector potential using Raman dressing~\cite{Lin2009a,Lin2009b}.  A (real) uniform magnetic field $\mathbf{B}_0 = B_0\ey$ Zeeman split adjacent $m_F$ levels by $E_{\rm Z}/h = 3.25 \MHz$.  Two Raman beams at $\lambda \approx 790.1$~nm  counterpropagated along $\pm\ex$ (with frequencies $\omega_{\rm R1,R2}$) and  coupled the three $m_F$ levels with a strength of $\hbar\Omega_{\rm R} = 5.6(6) E_{\rm L}$ for the TOF measurements and $\Omega_{\rm R} = 6.3(3) E_{\rm L}/\hbar$ for the \emph{in situ} measurements, where $E_{\rm L} = h^2/2m\lambda^2 = h \times 3.67 \kHz$ is the recoil energy.  The detuning $\hbar\delta = \hbar(\omega_{\rm R 1} - \omega_{\rm R 2})-E_{\rm Z}$ from Raman resonance was set to zero at the centre of the trap. To realize a synthetic magnetic field, we introduced a spatial dependence to the detuning with a (real) magnetic field gradient along $\ey$ such that $\mathbf{B}(\mathbf{r}) \approx \mathbf{B}_0 + b^\prime y  \ey$; values of $\mu_{\rm B}b^\prime/h$ range from 0 to 640 Hz/\textmu m in this experiment.  The resulting vector potential $\mathbf{A}^* = B^* y~ \ex$ gives a \emph{synthetic} magnetic field $\mathbf{B^*} = B^* \ez$, with a cyclotron frequency $\Omega_{\rm C} = q^*B^*/m^*$.  
The Raman dressing in this geometry altered the dispersion relationship only along $\ex$, giving an effective mass $m_x^*/m = 2.6(1)$.

\subsection*{Generating currents}
We modulated the confining potential energy $U(\mathbf{r},t) = U(\mathbf{r}) + \delta U_x(x) \sin (\omega t)$, with amplitude $\delta U_x(x) = \delta\kappa_x x^2/2$ and frequency $\omega$ to drive an ac current along $\ex$.  To allow the system to reach steady-state, we smoothly turned on the modulation amplitude in 500~ms from zero to $\delta \kappa_x/\kappa_x = 0.074 (2)$ 
for the TOF images, and 
to $\delta \kappa_x/\kappa_x = 0.151(3)$ 
for the \emph{in situ} images (a greater modulation amplitude was required to obtain an acceptable signal \emph{in situ}).
This larger modulation corresponded to a change of the \emph{in situ} size of the condensate of $\lesssim20 $\% peak-to-peak.  We verified that the drive was in the linear response regime with measurements at $\delta \kappa_x$ both less and greater than the values used here, and found that the response was proportional to $\delta \kappa_x$.

\subsection*{\emph{in situ} image analysis}
After modulating at full-strength for a time $t_{\rm mod}$, we prepared the system for imaging: in the final 0.5 ms during which the BEC remained trapped, the magnetic field gradient was removed and the magnetic bias field was aligned along  $\ez$ for imaging.  This process, rapid compared to the periods associated with trap frequencies, did not noticeably affect $n(x,y)$.  Upon removing the optical trapping potential, a weak 100 \textmu s optical pumping pulse, detuned $\approx 140\MHz$ from the $F=1$ to $F^\prime =2$ transition, transferred about 2~\% of the atoms (randomly sampled from the distribution) to the $F = 2$ hyperfine manifold.  These atoms were resonantly imaged in 40 \textmu s on the $\ket{F = 2, m_{F} =2} \rightarrow \ket{F^\prime = 3, m_{F}^\prime =3}$ cycling transition.  The peak optical density was generally about $1.5$.  

We obtained the second-order moments described in the text (equation (3)) 
from the 2D density profiles ${n}(x,y)$, shown in Fig.~2.  
We checked that the thermal (non-BEC) component of the cloud did not significantly contribute to the optical density signal in these images by using fits to the appropriate 2D atomic density distribution of a mixed (mean-field BEC $+$ thermal) trapped cloud.  Since our results were not affected (within experimental uncertainty) by including the thermal component, we opted to ignore it in favour of a simpler analysis.  We determined the region over which we calculated the second-order moments as follows: we fit $n(x,y) \sim [1-(x/R_x)^2-(y/R_y)^2]^{3/2}$, with a 2D gaussian, and analyzed those points within the area for which the gaussian's value was $\geq 1~\%$ of its peak.  Repeating this analysis with a $5~\%$ limit gave similar results, confirming that the $1~\%$ limit effectively captures the entire BEC.

For the \emph{in situ} data presented in Fig.~4, 
the absorption of the optical pumping pulse as it propagated through the cloud caused a small asymmetry to the observed $F = 2$ density profiles.  To ``smooth over'' this  asymmetry, we assume a  3D Thomas-Fermi density profile of the form:
\begin{align}
\label{eq:TF}
n(x,y,z,t) &= \frac{\mu}{g}\bigg\{1-\left[\frac{x}{R_x(t)}\right]^2\!- \left[\frac{y}{R_y(t)}\right]^2\!-\nonumber\\ &\qquad\qquad\qquad\left[\frac{z}{R_z(t)}\right]^2\!- s_{xy}(t)xy\bigg\},
\end{align}
where $\mu$ is the chemical potential; $R_{x,y,z}(t) = R_{x,y,z}(0) + \delta R_{x,y,z}(t)$ is the time-dependent Thomas-Fermi radius of the cloud along $\exyz$ with equilibrium radius $R_{x,y,z}(0)$; and $s_{xy}(t)$ is a skew parameter characterizing the Hall response, whose equilibrium value is zero.  Fits of equation \eqref{eq:TF}, integrated over $z$, to the observed density give radii $R_x$, $R_y$ and skew parameter $s_{xy}$.  Exact integration over the density profile (equation \eqref{eq:TF}) gives  the  second-order moments in terms of the Thomas-Fermi parameters: $\expect{x_i^2} = 4 R_i^2/\left[28-7\left(s_{xy} R_x R_y\right)^2\right]$ and $\expect{xy} = -2 s_{xy}R_x R_y/\left[28-7\left(s_{xy} R_x R_y\right)^2\right]$.  

Lock-in analysis is performed on the normalized second-order moments to find the amplitude response, as plotted.  We calculated the ``in-phase'' response by multiplying the  time-dependent signals $\expect{x^2}$, $\expect{y^2}$ and $\expect{xy}$ by $\sin(\omega t)$ and extracting the average of the product to give $\Delta\expect{x^2}$, $\Delta\expect{y^2}$ and $\Delta\expect{xy}$.  Similarly, we calculated the ``out-of-phase'' response by multiplying by $\cos(\omega t)$.  The displayed uncertainties in these values  represent the standard deviation of these means.  We calculated the amplitude response of each moment  as the root-sum-square of the in- and out-of-phase responses.

\subsection{TOF image analysis}
After modulating at full-strength for a time $t_{\rm mod}$, we suddenly removed the optical trap (in $\lesssim 1\us$).  During the following 2~ms of TOF, we adiabatically swept the Raman detuning $\delta$ from 0 to $-77~E_{\rm L}$, adiabatically transferring (deloading) the Raman-dressed spin-momentum superposition state to $\ket{F = 1, m_F = +1}$.  During the sweep, the subsequent turn-off of the Raman beams, and the remainder of TOF, the condensate underwent free expansion, initially driven by the mean field.  

We absorption-imaged the expanded atomic distribution along $\ez$ after 36.2~ms TOF.  From the resulting two-dimensional density profile ${n}(x,y)$, we determined the second-order moments described by equation (3).  
The discrete sums were analyzed over a region defined as described above.  To account only for the dynamics in the condensate, we removed the  ($30~\%$ to $50~\%$) thermal background from the images by means of a gaussian fit to the region outside the condensate.  We extracted the lock-in response as above.

\end{document}